\documentstyle[eqsecnum,preprint,aps,epsbox]{revtex}

\def\beq{\begin{equation}}
\def\eeq{\end{equation}}
\def\bx{{\bf x}}
\def\bk{{\bf k}}
\newcommand{\mpl}{{m_{\rm Pl}}}
\newcommand{\etal}{{\it et al.\ }}
\renewcommand{\prd}{{\it Phys.\ Rev.\ D~}}
\renewcommand{\prl}{{\it Phys.\ Rev.\ Lett.\ }}
\newcommand{\plb}{{\it Phys.\ Lett.\ B~}}

\begin{document}
 
\draft
\title{Generality of Topological Inflation}

\author{Nobuyuki Sakai\thanks{Electronic address: 
sakai@ke-sci.kj.yamagata-u.ac.jp}}
\address{Faculty of Education, Yamagata University, Yamagata 
990-8560, Japan}
\address{Graduate School of Science and Engineering, Yamagata 
University, Yamagata 990-8560, Japan}
\address{Osservatorio Astronomico di Roma, Via Frascati, 00040 Monte 
Porzio Catone, Italia}

\maketitle

\begin{abstract}
Many authors claimed that a large initial inhomogeneity prevents the onset 
of inflation and therefore inflation takes place only if the scalar field 
is homogeneous or appropriately chosen over the horizon scale. We show 
that those arguments do not apply to topological inflation. The core of a 
defect starts inflation even if it has much smaller size than the horizon 
and much larger gradient energy than the potential, as long as the 
vacuum expectation value is large enough ($\gtrsim0.3\mpl$) and the core is not 
contracting initially. This is due to stability of false vacuum.
\end{abstract}

\vskip 1cm
\begin{center}
PACS number(s): 98.80.Cq, 11.27.+d, 04.20.Dm
\end{center}

\newpage
\tighten
\section{Introduction}

Inflation gives a natural solution of the horizon problem of the big-bang 
universe. However, if inflation requires homogeneous initial conditions 
over the super-horizon scale, one cannot say that it is a solution of the 
horizon problem, though it reduces the problem by many orders of 
magnitude. To answer this naive question, many people have investigated 
how initial inhomogeneity affects the onset of inflation 
\cite{GP,KK,CNN,confGP,VT,Cou,BG}. The most remarkable work was done by 
Goldwirth and Piran \cite{GP}, who solved the full Einstein equations for 
a spherically symmetric  spacetime. Their results are summarized as 
follows. New inflation is so sensitive to initial inhomogeneity that it 
requires homogeneity over a region of several horizon sizes is needed. 
Chaotic inflation is not so affected by initial inhomogeneity but requires 
a sufficiently high average value of the scalar field over a region of 
several horizon sizes. Later many people explored the same problem in 
different approaches \cite{CNN,confGP,VT,Cou,BG}, which as a whole confirmed 
the results obtained by Goldwirth and Piran. This suggests failure of the 
``ordinary" scenario of inflationary cosmology and therefore some 
alternatives were discussed \cite{VT,Cou,BG},

In this paper we investigate whether topological inflation \cite{LV} also
requires initial homogeneity over the horizon size. The simple model of 
topological inflation is described by the potential,
\beq\label{V}
V={\lambda\over 4}(\Psi^2-\eta^2)^2.
\eeq
However, the same type of inflation takes place in many plausible theories 
such as a nonminimally coupled massive scalar field \cite{SY}, some models of 
supergravity \cite{sugra}, and some superstring inspired models 
\cite{string,BAR}. 
The inflationary condition for the model (\ref{V}) was investigated by 
numerical analysis \cite{SSTM}: $\eta\gtrsim0.33\mpl$ for non-gauged defects. 
Although we claimed that the critical value of $\eta$ does not depend on 
initial conditions \cite{SSTM}, it seems to lack of convincible 
explanations according to recent arguments \cite{VT,Cou,BG}. We therefore 
return to the same model and clarify the effect of initial inhomogeneity 
both analytically and numerically.

The plan of this paper is as follows. In Sec II, we discuss analytically the 
scalar field dynamics under the approximation that the metric is homogeneous.
In Sec III, we numerically solve the Einstein equations for a global monopole
with a large gradient. Section IV is devoted to a summary and discussion.
We use the units of $c=\hbar=1$ throughout the paper.

\section{Scalar Field Dynamics in a Homogeneous Spacetime}

Let us consider the scalar field dynamics under the approximation that 
the metric is homogeneous, following Berera and Gordon \cite{BG}.
The equation of the scalar field $\Psi(t, \bx)$ is
\beq\label{sfeq}
\ddot\Psi+3H\dot\Psi-{\nabla^2\Psi\over a^2}+{dV\over d\Psi}=0,
\eeq
where an overdot denotes time-derivatives, $a(t)$ is the scale factor and 
$H\equiv\dot a/a$.
Although $a(t)$ is not a solution of the exact Einstein equations, it is 
instructive to write down its evolution equation under the present 
approximation,
\beq\label{aeq}
{\ddot a\over a}={8\pi\over 3\mpl^2}(-\dot\Psi^2+V).
\eeq
This equation tells us that inflation does not take place ($\ddot a<0$) if 
the kinetic term dominates over the potential. It should be noted that a gradient
term does not appear in Eq.(\ref{aeq}). Gradient energy can affect cosmic expansion
only by exciting kinetic energy through the field equation (\ref{sfeq}).

Because inflation can take place at $\Psi\approx0$, we approximate the 
potential (\ref{V}) as $V=\lambda\eta^4/4-\lambda\eta^2\Psi^2/2$. By 
Fourier expansion
$\Psi(t,\bx)=\sum\Psi_{\bk}(t)e^{i\bk\cdot\bx}$, 
Eq.(\ref{sfeq}) becomes
\beq\label{keq}
\ddot\Psi_{\bk}+3H\ddot\Psi_{\bk}+{k^2\over a^2}\Psi_{\bk}-\lambda\eta^2 
\Psi_{\bk}=0.
\eeq
We assume the slow-roll condition for a homogeneous and isotropic 
spacetime, $|\ddot\Psi|\gg3H|\dot\Psi|$ and $\dot\Psi ^2/2\gg V$ at 
$\Psi\approx0$, which is equivalent to $\eta\gg\mpl/\sqrt{6\pi}$. 
(Actually the numerical finding $\eta>0.33\mpl$ \cite{SSTM} is more precise 
condition.) Under 
this condition what happens if the $k^2$-term of (\ref{keq}) dominates? 
Clearly $\Psi_{\bk}$ begins oscillation, which contributes kinetic energy. 
In the case of other nontopological inflationary models, inflation never 
happens if kinetic energy dominates until the mean field rolls down to the 
potential minimum and the slow-roll condition breaks down. On the other 
hand, in the case of topological inflation, the false vacuum $\Psi=0$ is 
topologically stable and never decays. As long as the core region 
$\Psi\approx0$ is not contracting at an initial time ($\dot a\ge 0$), 
oscillations are eventually damped. Because a local region of $\Psi\approx 
0$ always exists, inflation takes place eventually.

The above arguments are rough because we have neglected the inhomogeneity of
the spacetime. This is why we shall move on to numerical analysis in the next
section. Nevertheless, the above arguments will help us to understand our
numerical results intuitively.

\section{Numerical Analysis of a Monopole with a Large Gradient}

Here we take inhomogeneity of spacetime into account by solving the 
Einstein equations coupled to the scalar field equation numerically. We 
restrict ourselves to the global monopole system. We believe that the main 
features of dynamics do not change for domain walls and global strings, 
according to the previous work \cite{SSTM}. Because we have already known 
that the inhomogeneity of gauge fields affects the onset of inflation 
\cite{Sak} and our present concern is pure effects of the scalar field 
inhomogeneity, we do not consider gauged defects here.

We therefore consider the Einstein-Higgs system, which is described by the
action,
\beq\label{action}
{\cal S}=\int d^4x\sqrt{-g}\left[{\mpl^2\over 16\pi}{\cal R}
-\frac12(\partial_{\mu}\Psi^a)^2-V(\Psi) \right], ~ ~ (a=1,2,3),
\eeq
with the potential (\ref{V}). 
The action (\ref{action}) yields the field equations,
\begin{eqnarray} \label{Ein}
G_{\mu\nu}&\equiv&{\cal R}_{\mu\nu}-\frac12g_{\mu\nu}{\cal R}
={8\pi\over\mpl}T_{\mu\nu},\\
\label{Psi}
\Box\Psi^a&=&\frac{\partial V(\Psi)}{\partial\Psi^a},
\end{eqnarray}
with
\begin{equation}
T_{\mu\nu}\equiv\nabla_{\mu}\Psi^a\nabla_{\nu}\Psi^a
-g_{\mu\nu}\left[\frac12(\nabla\Psi^a)^2+V(\Psi)\right].
\end{equation}
We assume a spherically symmetric spacetime 
and adopt the metric,
\beq\label{metric}
ds^2=-dt^2+A^2(t,r)dr^2+B^2(t,r)r^2(d\theta^2+\sin^2\theta d\varphi^2).
\eeq
For the scalar field, we adopt the hedgehog ansatz:
\beq\label{hg}
\Psi^a=\Psi(t,r)(\sin\theta\cos\varphi,\sin\theta\sin\varphi,\cos\theta)
\eeq

In the following we shall write down the field equations (\ref{Ein}) and
(\ref{Psi}) with the assumptions (\ref{metric}) and (\ref{hg}).
To begin with, we introduce the extrinsic curvature tensor $K_{ij}$:
\begin{equation}
K^r_r=-{\dot A\over A},~~~ K^{\theta}_{\theta}(=K^{\varphi}_{\varphi})
=-{\dot B\over B},~~~
K\equiv K^i_i.
\end{equation}
Next, following Nakamura {\it et al.}\cite{Nak}, we introduce 
auxiliary dynamical variables,
\begin{equation}
a\equiv{A-B\over r^2},~~~ k\equiv{K^{\theta}_{\theta}-K^r_r\over r^2},
\end{equation}
and use a new space variable, $x\equiv r^2$, instead of $r$. The advantages of
those variables were discussed in Ref.\cite{SYM}.

We introduce further auxiliary variables,
\beq
C\equiv-{B'\over B}, ~~~
\psi\equiv{\Psi\over r}, ~~~ \varpi\equiv\dot\psi, ~~~ \xi\equiv\psi',
\eeq
where a prime denotes $\partial/\partial x$. 
A full set of dynamical variables is 
$a,~B,~C, ~\psi,~\xi, ~K,~k$ and $\varpi$. The time-derivatives of the first
five variables are given by the definitions above:
\begin{eqnarray}
&&\dot a=-a K^r_r+B k,~~~
\dot B=-B K^{\theta}_{\theta}, ~~~ \dot C={K^{\theta}_{\theta}}',
\label{dotB} \\ \label{dotpsi}
&&\dot\psi=\varpi, ~~~ \dot\xi=\varpi'.
\end{eqnarray}
Note that the value of ${K^{\theta}_{\theta}}'$ in Eq.(\ref{dotB}) is
determined by the momentum constraint (\ref{MC}) below, which gives a more 
accurate value than a finite difference of $K^{\theta}_{\theta}$.

Now we write down the field equations (\ref{Ein}) and (\ref{Psi}) as
\begin{eqnarray}\label{HC}
-G^t_t&\equiv&{K^2-k^2x^2\over 3}
+{8x\over A^2}\left(C'-{A'C\over A}-{3C^2\over 2}\right)
+{4\over A^2}\left[{A'\over A}+4C+{a(A+B)\over 4B^2}\right] \nonumber\\
&=& {8\pi\over\mpl^2}\left[{x\varpi^2\over 2}+{2x\xi\over A^2}(x\xi+\psi)
+{\psi^2\over 2A^2}+{\psi^2\over B^2}+V \right],
\\ \label{MC}
{G_{tr}\over 4r}&\equiv& {K^{\theta}_{\theta}}'+k\left({1\over2}-xC\right)
={2\pi\varpi\over\mpl^2}(2x\xi+\psi),\\
\label{dotK}
-{\cal R}^t_t&\equiv&\dot K-{K^2+2x^2k^2\over 3}
={8\pi\over\mpl^2} (x\varpi^2-V),\\
\label{dotk}
{{\cal R}^r_r-{\cal R}^{\theta}_{\theta}\over r^2}&\equiv&
\dot k-kK+{2\over A^3}\left[2AC'-2A'C+aC+a'-{a^2\over B^2}
\left({A\over 2}+B\right)\right] \nonumber\\
&=&{8\pi\over\mpl^2 A^2}
\left[4\xi(x\xi+\psi)-{\psi^2 a(A+B)\over B^2}\right],\\
\label{dotvarpi}
\dot\varpi&=&K\varpi+{4\over A^2}\left[x\xi'-x\xi\left({A'\over A}+2C\right)
+{5\xi\over 2} \right] \nonumber\\
&&-{2\psi\over A^2}\left[{A'\over A}+2C+{a(A+B)\over B^2}\right] 
-\lambda\psi(\Psi^2-\eta^2).
\end{eqnarray}
Equations (\ref{HC}) and (\ref{MC}) are constraint equations, which fixes
initial values of the metric. Equations (\ref{dotK})-(\ref{dotvarpi}) together 
with (\ref{dotB}) and (\ref{dotpsi}) provides the time evolution of the seven 
variables.

Let us discuss how to discretize
a space at each time into a mesh and approximate spatial derivatives.
Because our field equations are written with $x$ instead of $r$, the simplest
way is to use a regular mesh with separation $\Delta x$ in $x$-space, as we did in
Ref.\cite{SYM}.
However, this mesh gives lower resolution for a center than for a far region in
terms of $r$-space, which is unsuitable for the present case, where the central 
region has a large gradient. Therefore, we adopt a regular mesh with separation 
$\Delta r$ in $r$-space,
\beq
r_1=0,~ r_2=\Delta r,~ ...,~ r_i=(i-1)\Delta r,~ ...,
\eeq
and approximate $x$-derivatives of any function $F(x)$ by
\beq
{dF\over dx}(x_i)={1\over 2r_i}{dF\over dr}(x_i)
\approx{1\over 2r_i}{F_{i+1}-F_{i-1}\over2\Delta r}
={F_{i+1}-F_{i-1}\over x_{i+1}-x_{i-1}}.
\eeq

We impose further regularity condition on all dynamical variables at $r=0$, 
$F(0)$, as follows. Although all evolution equations do not contain any 
diverging factor at $r=0$ like $1/r$, we do not use them to obtain $F_1=F(0)$
at each time. Instead, following Hayley and Choptuik \cite{HC}, we 
employ a ``quadratic fit",
\beq
F_1={4F_2-F_3\over 3}.
\eeq
We find that this simple method is very effective to keep good accuracy in the
center.

As initial conditions of the scalar field, we suppose
\beq
\Psi(t=0,r)=\eta\tanh\left({r\over\delta}\right), ~~~
\dot\Psi(t=0,r)=0,
\eeq
where $\delta$ is a free parameter, which controls the initial gradient at 
$\Psi\approx0$. If the previous arguments applied to topological 
inflation, $\delta$ would have to be larger than the horizon size 
$H_{0}^{-1}\equiv(8\pi V(0)/3\mpl^2)^{-1/2}$ for the onset of inflation. 
Now our question is specified: can inflation start even if $\delta \ll 
H_{0}^{-1}$? We therefore choose $\delta=0.01H_0^{-1}$ below as an example.

For the metric, there are four unknown variables, $B,~a,~K$ and $k$, 
while there are two constraint equations (\ref{HC}) and (\ref{MC}). 
Therefore, two of the four variables are arbitrary chosen, and this choice is 
 an important problem. If we choose $K=$constant, for example, 
large initial inhomogeneity of $\Psi$ causes a black hole or a 
singularity, as discussed by Chiba \etal \cite{CNN}~~
The appearance of black holes or singularities, however, does not mean failure 
of inflation, because inflation is a local phenomenon.
To judge whether inflation occurs or not, 
we should solve the evolution equations; however, it is difficult to solve them 
for such spacetimes numerically. For such a technical reason, we set initial 
conditions without a black hole nor a singularity by adopting the 
following two ways, under the condition that the initial 3-space is not 
contracting, $K\le 0$.

(A) Following Goldwirth and Piran \cite{GP}, we set a homogeneous and 
isotropic 3-space by introducing another massless scalar field $\chi$ to 
compensate inhomogeneous energy density of $\Psi$. We assume 
$\chi(t=0,r)=0$ and $\dot\chi(t=0,r)>0$. The initial metric functions become
$B=1,~a=0,~k=0$ and $K=$constant. 

(B) Without introducing a massless scalar field, we set $B=1$ and $a=0$ and solve 
the constraint equations (\ref{HC}) and (\ref{MC}) to obtain $K$ and $k$.

We demonstrate solutions with $\lambda=0.1$, $\eta=\mpl$ and 
$\delta=0.01H_0^{-1}$ in Figs.\ 1 and 2. In both cases we see in (a) and (b) the
monopole core expands. As Eq. (\ref{dotK}) suggests, whether the expansion rate is
increasing or not is determined by the sign of $\dot\Psi^2-V$ even in an inhomogeneous
spacetime. Therefore, once kinetic energy becomes
smaller than the potential, inflation starts even though gradient energy is still
dominant. This behavior is actually seen in Figs.\ 1(c)-(e) and 2(c)-(d).

The evolution time in Fig.\ 2 is relatively short because calculation breaks down
soon after the black-hole horizon appear at $t\approx 0.27H_0^{-1}$. However, we
believe it demonstrates sufficiently that inflation actually begins.

It should be emphasized that topological inflation can take place even if  
$\rho_{grad}\gg V$, as shown in Figs.\ 1(c) and 2(c). This is in contrast 
with the previous results \cite{GP}, which showed that new inflation 
requires a more stringent condition than $\rho_{grad}<V$.

\section{Summary and Discussion}

We have investigated whether topological inflation requires 
initial homogeneity over the horizon size. Contrary to the results for 
nontopological inflationary models, which have been investigated by many 
people, topological inflation can take place even if the initial false 
vacuum has one percent of the horizon size and gradient energy of $10^5 
V(0)$. This is due to topological stability of false vacuum. In this sense 
topological inflation explains most naturally the onset of inflation in a 
really chaotic universe.

Here we make a comment on the papers by Vachaspati and Trodden \cite{VT}, who
considered initial conditions for inflation based on the null Raychaudhuri 
equation, and concluded that {\it homogeneity on super-horizon scales must be
assumed as an initial condition.} They claimed that the size of the initial
inflationary patch, $X$, must be greater than the inflationary horizon,
$H_{inf}^{-1}$, which must be larger than the (pre-inflationary) background
FRW inverse Hubble size at the time inflation starts, $H_{FRW}^{-1}$, that is,
\beq
X\ge H_{inf}^{-1}\ge H_{FRW}^{-1}.
\eeq
However, what they really proved is only the second inequality, 
$H_{inf}^{-1}\ge H_{FRW}^{-1}$, to which we have no objection.
On the other, they just {\it assumes} the first inequality, although its 
validity is more crucial. Actually, what we showed in the present paper is 
topological inflation can takes place even if $X<H_{inf}^{-1}$. 

Another note is added.
Stability of false vacuum is a key feature of 
topological inflation, which sometimes leads to the idea that inflation 
cannot end \cite{BAR}. It is, however, not to worry. Once inflation 
begins, the evolution of a local region is described by Eq.(\ref{sfeq}) 
without the gradient term. Therefore, just like new inflation or chaotic 
inflation, $\Psi$ simply rolls down to the potential minimum except for 
$\Psi=0$, that is, inflation ends in any region except the one point. More 
detailed arguments are given in Ref.\cite{SNH}.

\acknowledgements
The author thanks Luca Amendola for discussions and his hospitality at 
Osservatorio Astronomico di Roma. He also thanks David Coule for critical 
comments. Numerical computation of this work was carried out at the Yukawa 
Institute computer facility.
This work was supported in part by the exchanging-researcher project 
between Japan Society for the Promotion of Science and National Research 
Council of Italy, 2003, and by Ministry of Education, Culture, Sports, 
Science and Technology, Grant-in-Aid for Young Scientists (B), 2003, 
No.\ 00267402.


\newpage
\small

\begin{figure}
 \begin{center}
  (a)\psbox[scale=0.40]{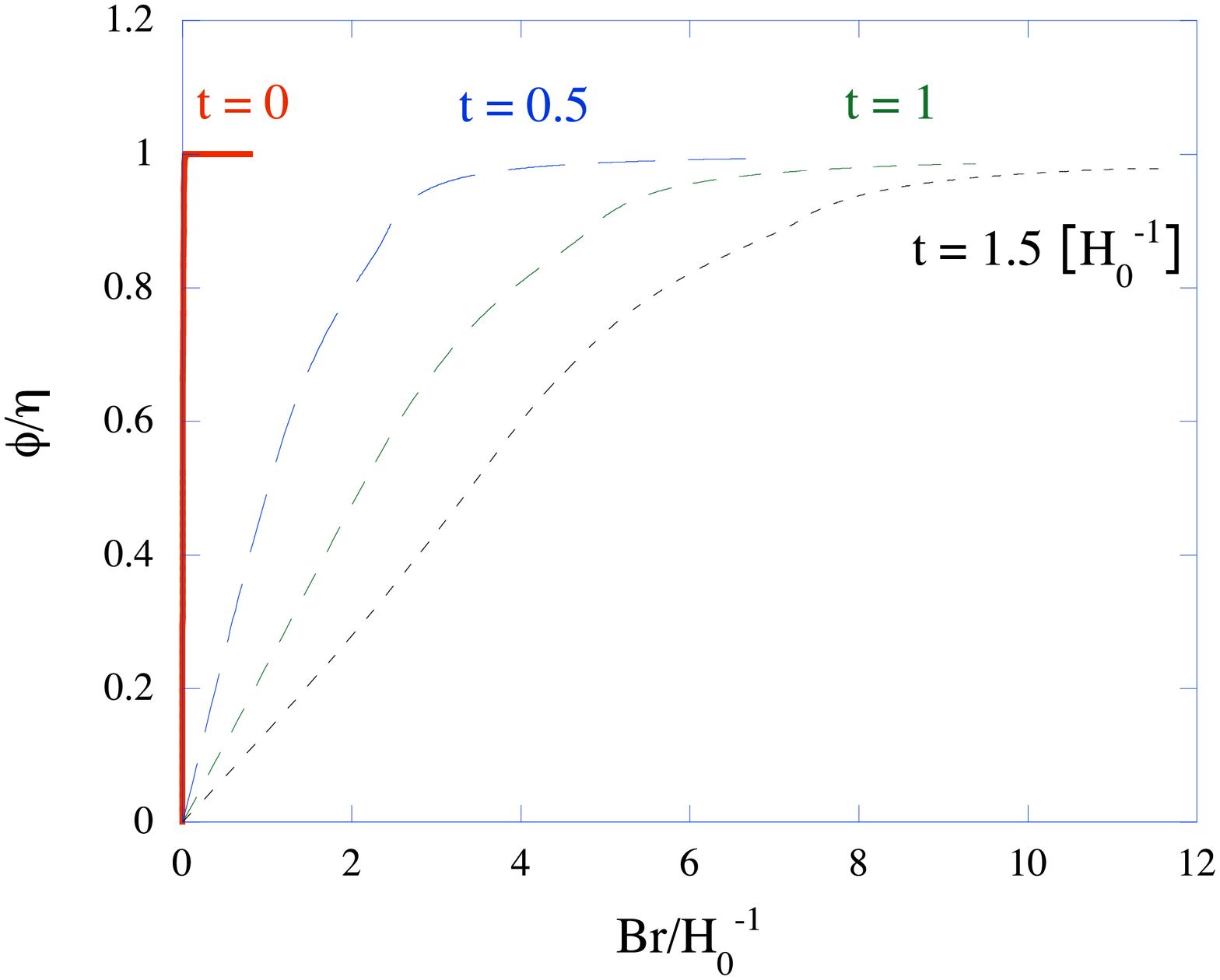}
  (b)\psbox[scale=0.40]{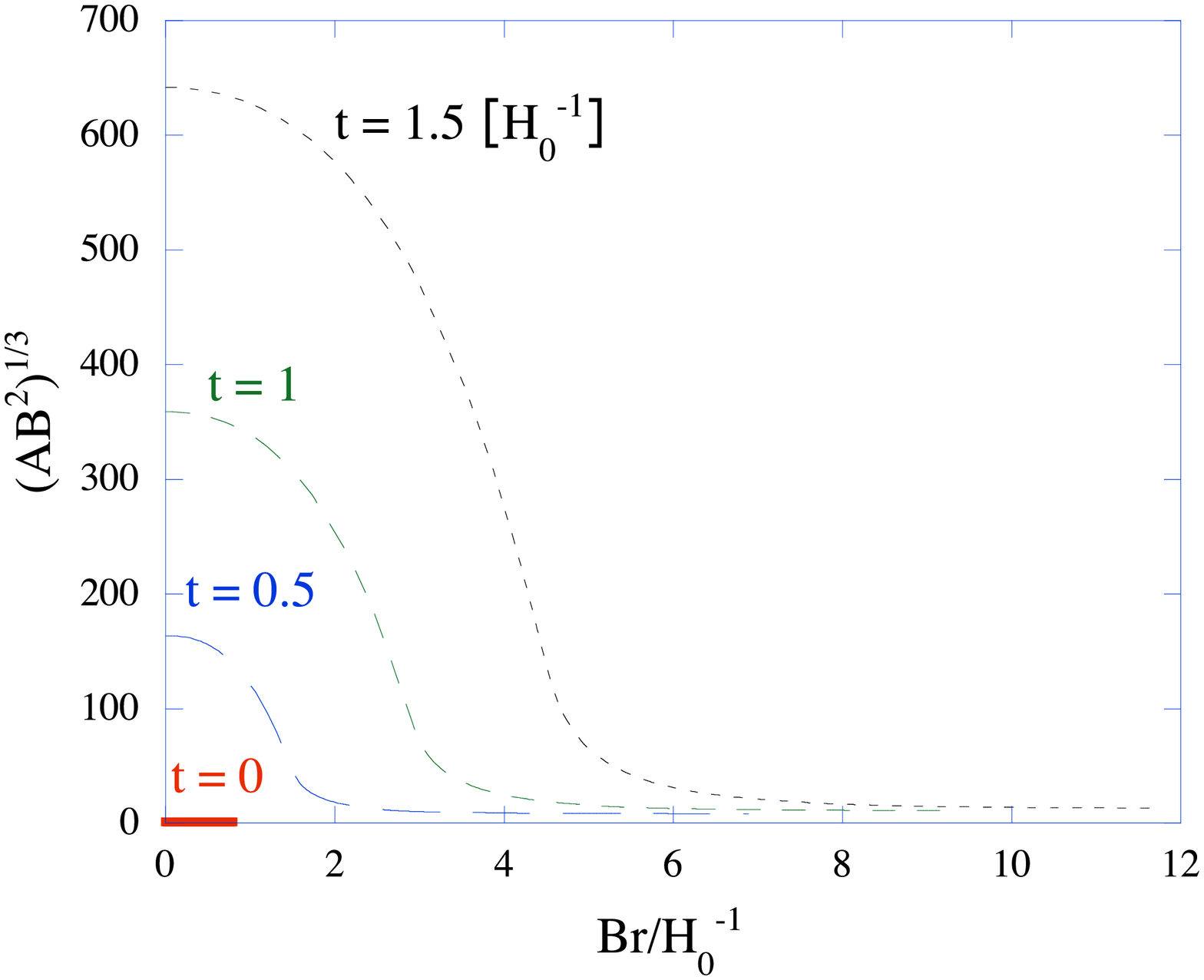}
  (c)\psbox[scale=0.40]{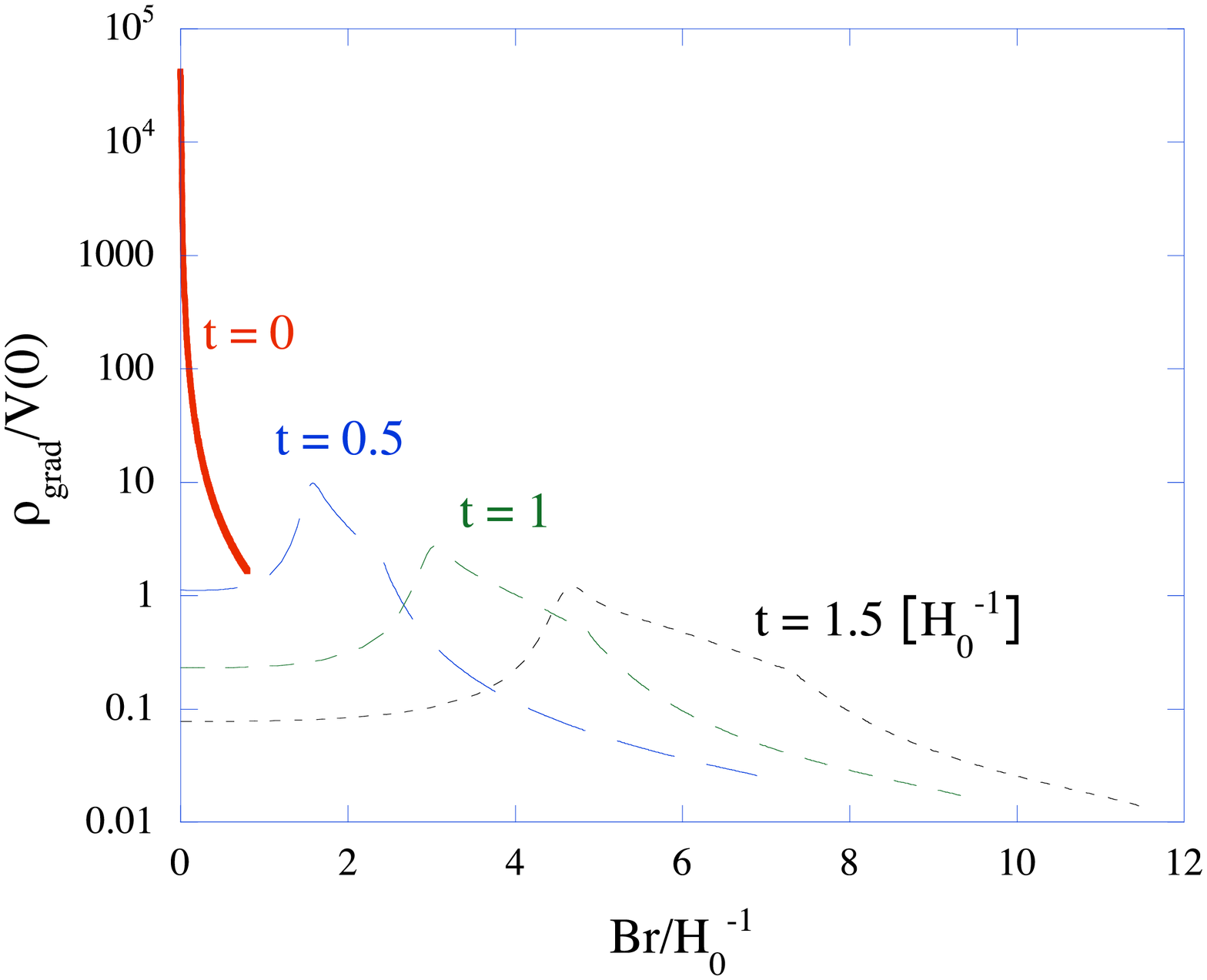}
\end{center}
\end{figure}

\newpage
\begin{figure}
 \begin{center}
  (d)\psbox[scale=0.40]{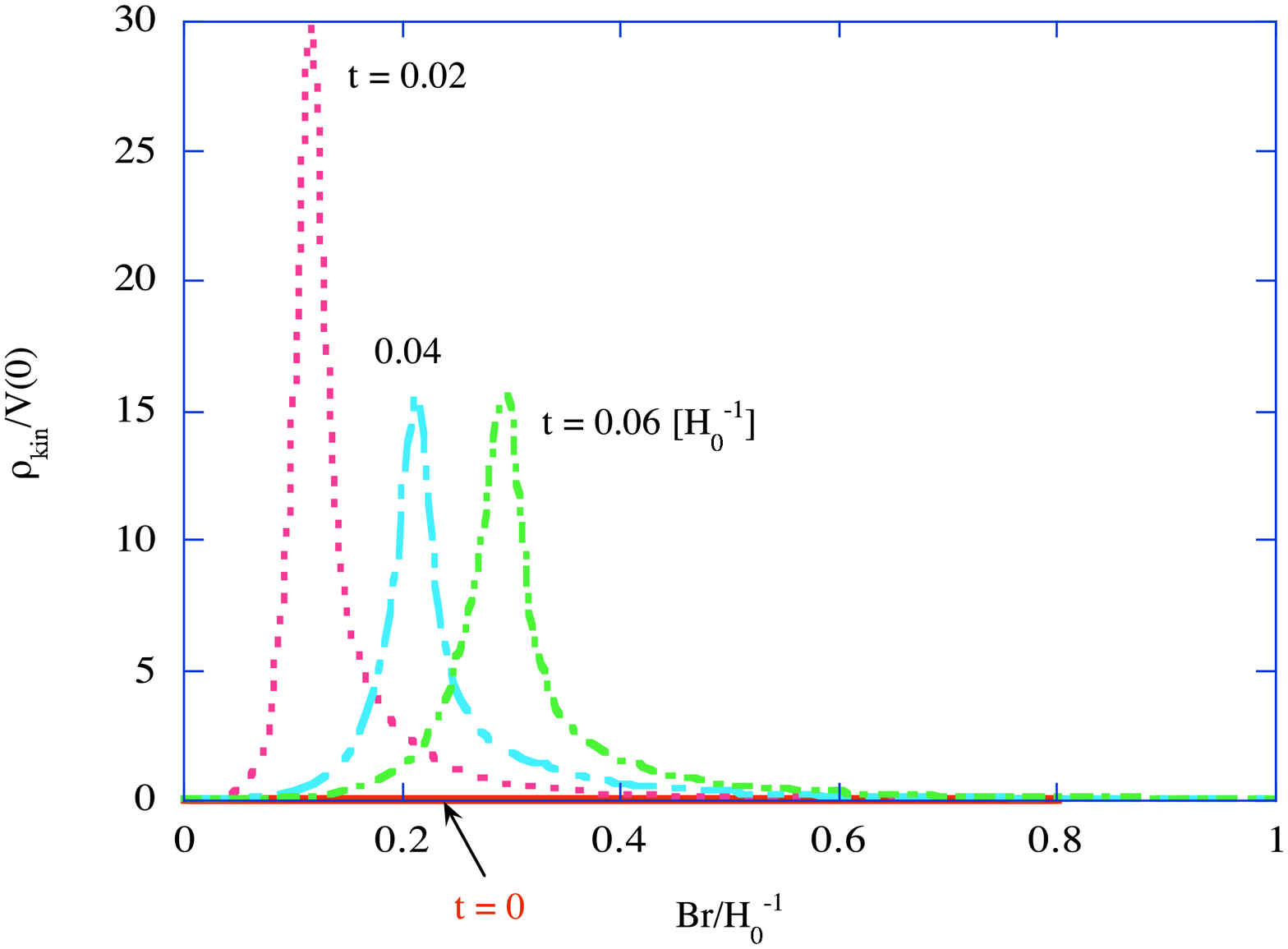}
  (e)\psbox[scale=0.40]{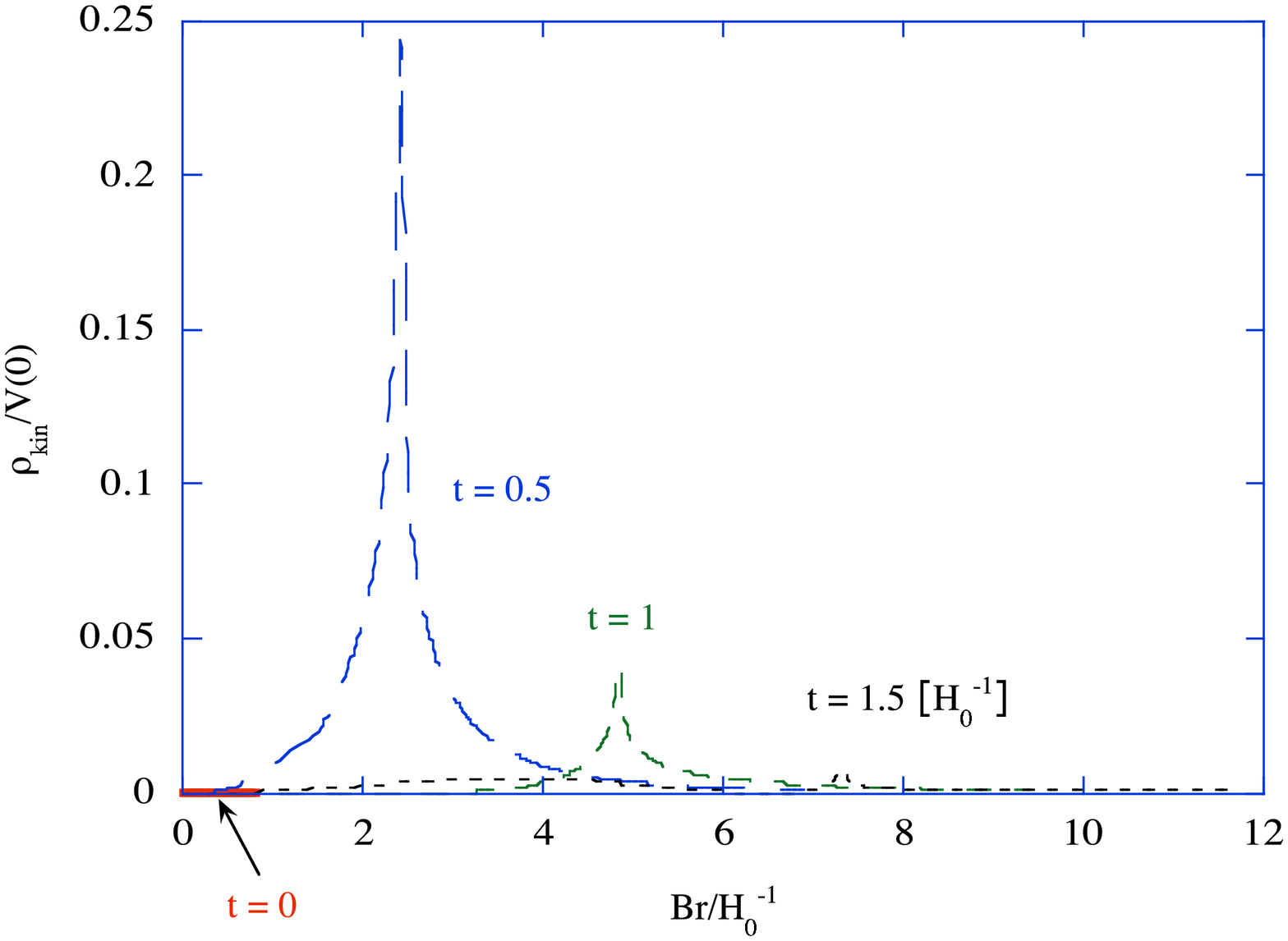}
\end{center}
\end{figure}

\noindent
{\bf FIG. 1}. A solution for the case (A). We show the evolution of the 
inflaton field, the volume element, and gradient energy
$\rho_{grad}=(\partial\Psi/\partial r)^2/(2A^2)$ in (a), (b) and (c), 
respectively.
For kinetic energy $\rho_{kin}=(\partial\Psi/\partial t)^2/2$, (d) and (e) 
report its early and late behaviors, respectively.
The abscissa is the arial radius of the spacetime $B(t,r)r$, normalized by the horizon size $H_0^{-1}$.

\begin{figure}
 \begin{center}
  (a)\psbox[scale=0.40]{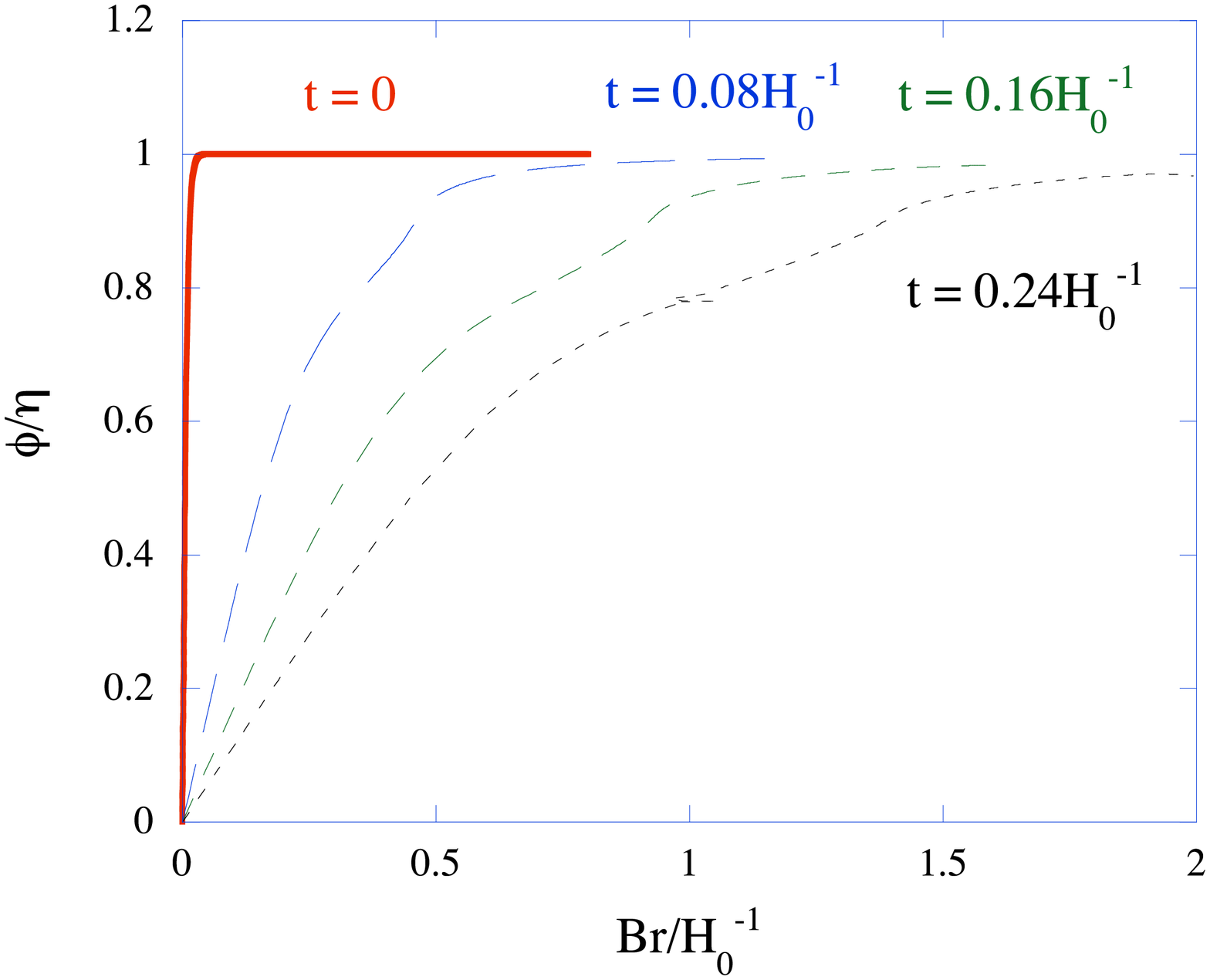}
\end{center}
\end{figure}

\newpage
\begin{figure}
 \begin{center}
  (b)\psbox[scale=0.40]{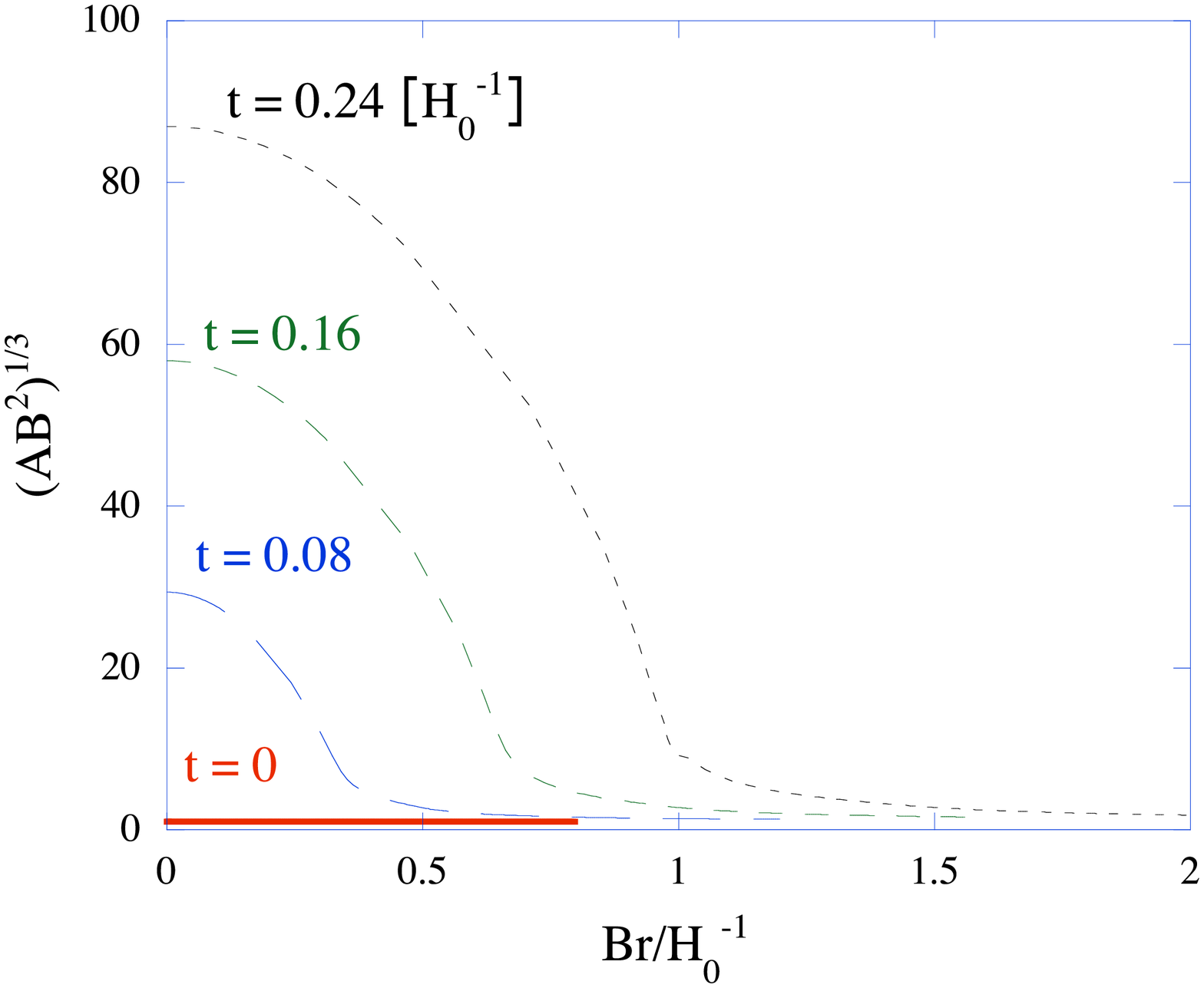}
  (c)\psbox[scale=0.40]{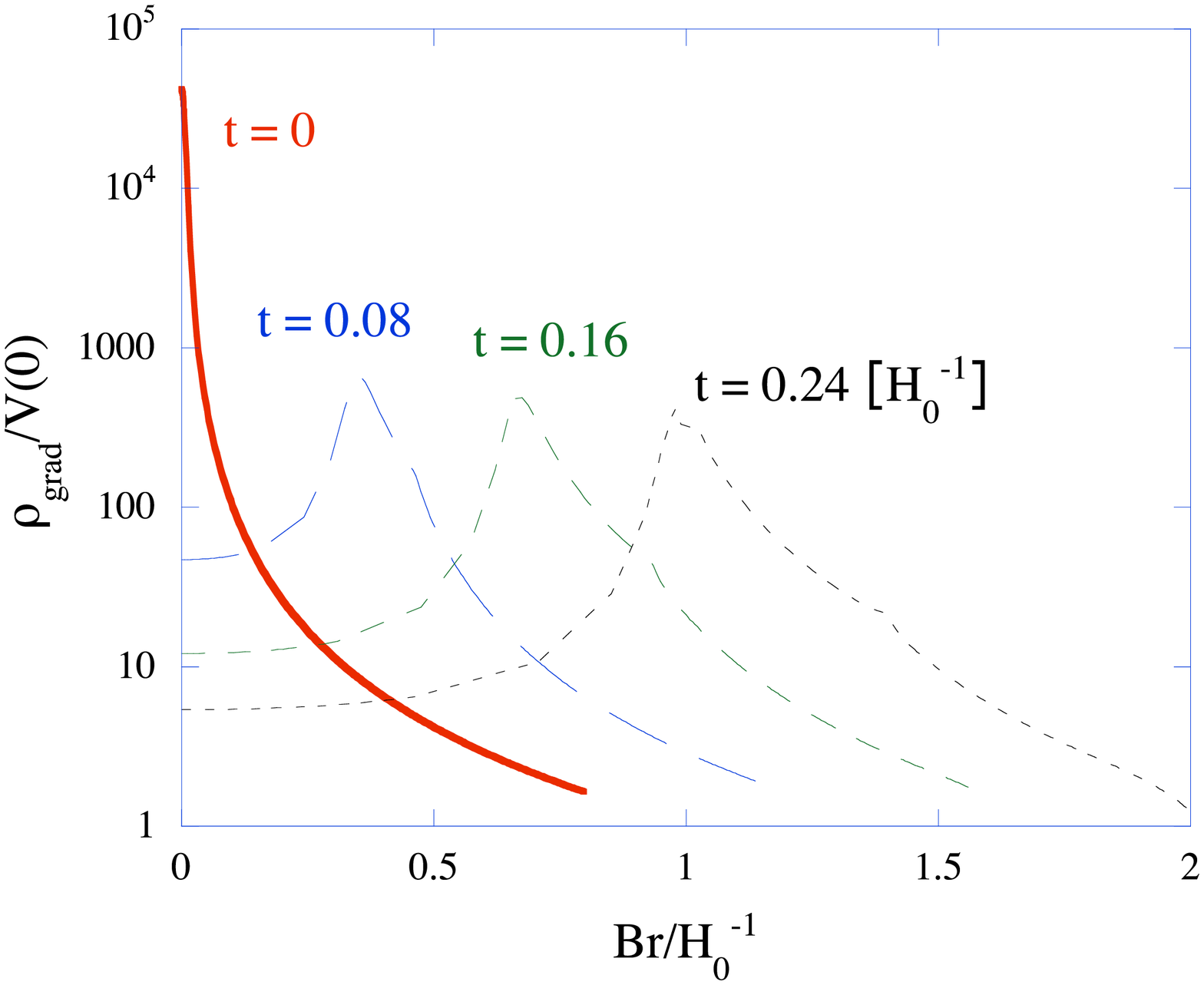}
  (d)\psbox[scale=0.40]{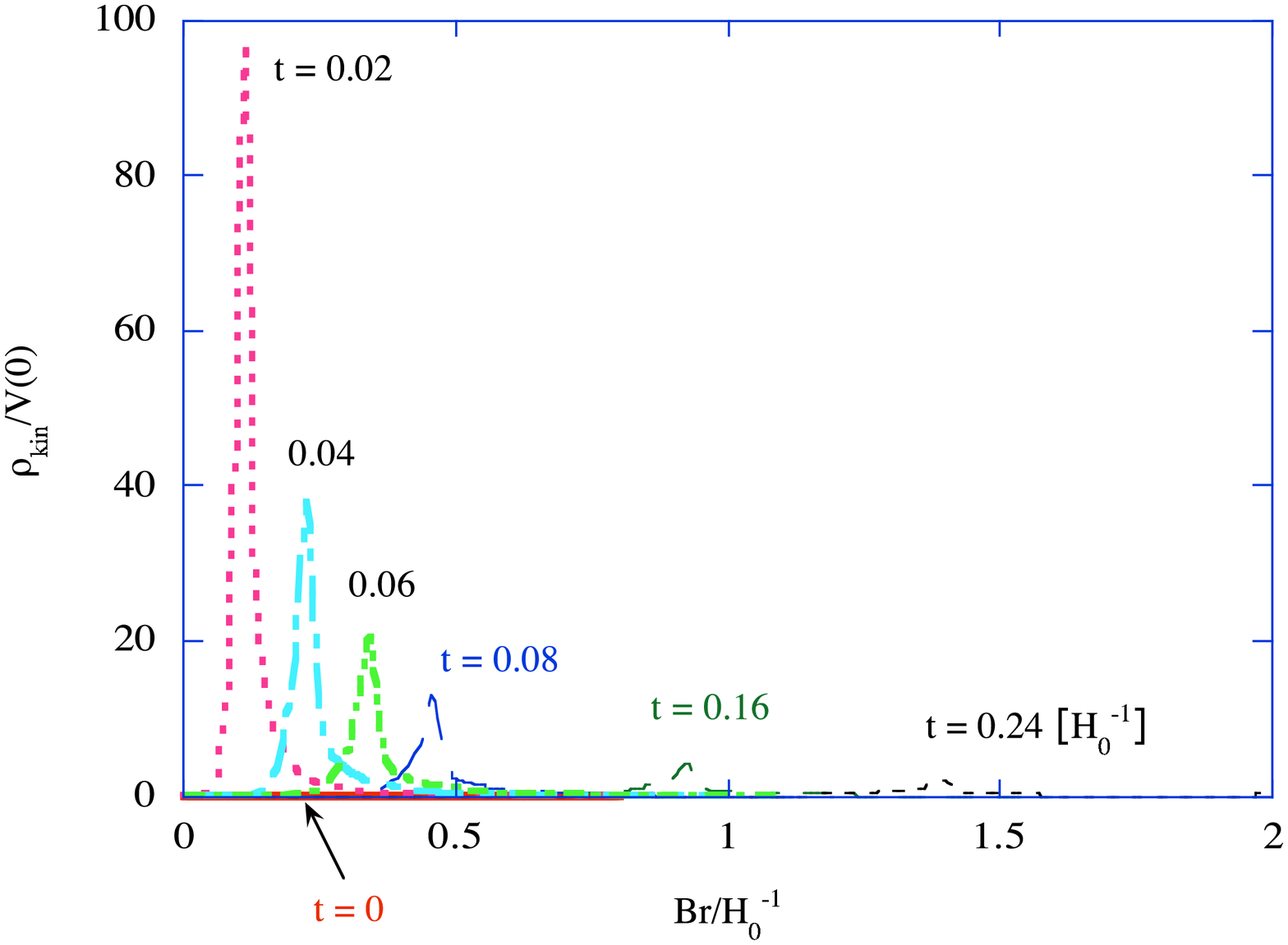}
\end{center}
\end{figure}

\noindent
{\bf FIG. 2}. A solution for the case (B). We show the evolution of the 
inflaton field, the volume element, gradient energy and kinetic energy in 
(a), (b), (c) and (d), respectively. 

\end{document}